# THERMAL CONDUCTIVITY OF METHANE-HYDRATE

A. I. Krivchikov, B. Ya. Gorodilov, O. A. Korolyuk, V. G. Manzhelii,
H. Conrad[*], W. Press[**]

*B.Verkin Institute for Low Temperature Physics and Engineering of NAN Ukraine, 61103, 47 Lenin Ave., Kharkov, Ukraine*
[*] *Institut für Festkörperforschung, Forschungszentrum Jülich GmbH 52425 Jülich, Germany*
[**] *Institut Laue-Langevin, Grenoble, France*

*The thermal conductivity of the methane hydrate $CH_4$ (5.75 $H_2O$) was measured in the interval 2÷140 K using the steady-state technique. The thermal conductivity corresponding to a homogeneous substance was calculated from the measured effective thermal conductivity obtained in the experiment. The temperature dependence of the thermal conductivity is typical for the thermal conductivity of amorphous solids. It is shown that after separation of the hydrate into ice and methane, at 240 K, the thermal conductivity of the ice exhibits a dependence typical of heavily deformed fine-grain polycrystal. The reason for the glass-like behavior in the thermal conductivity of clathrate compounds has been discussed. The experimental results can be interpreted within the phenomenological soft-potential model with two fitting parameters.*

## 1. Introduction

The methane clathrate hydrate is a compound whose crystal lattice contain 85% of water molecules forming a polyhedral clathrate structure. Solitary $CH_4$ molecules reside inside the clathrate cavities. The weak Van der Waals and electrostatic interactions between the guest molecule and its crystal environment stabilize the cubic structure of the hydrate and binds the local low-frequency vibrations of the guest molecules to the phonon modes of the matrix.[1,2] Nevertheless, most of physical properties of methane hydrate and other gas hydrates are similar to those of the ordinary ice Ih. The only exception is the extraordinary low thermal conductivity of $CH_4$ hydrate. Gas hydrates can be taken as model systems to investigate and clarify the origin of the glass-like behavior of thermal conductivity in crystal-structure solids.

The aim of the measurements is to investigate the heat transport phenomena in simple gas hydrates. Information on the phonon scattering mechanisms in gas hydrates can be derived from their thermal conductivity data. The dynamic interaction of the quantum rotor $CH_4$ with its crystalline environment is studied by measuring of the thermal conductivity of the methane hydrate samples. It is shown[3] by incoherent inelastic neutron scattering that the translational motion of these molecules is strongly coupled to the vibrations of the host lattice and is heavily damped at higher temperatures. Experiment aims at the understanding scattering mechanisms at the origin of the low thermal conductivity. For this purpose low temperature experiments, as performed here, are needed.

## 2. The sample preparation and method of measuring of their thermal conductivity

Methane hydrate $CH_4(5.75H_2O)$ sample was prepared at the Göttinger Zentrum Geowissenschaften, Germany. Methane clathrate was produced in a high pressure reaction chamber at 6 MPa and a temperature in the vicinity of 273 K. The starting material (ice Ih powder) consisted of spherical grains of a few tens to hundreds of μm in diameter. The clathration reaction lasted for 3 weeks. The procedure of the formation of porous gas hydrates from ice powders are detailed elsewhere.[4] The porous $CH_4$ hydrate crystallites sintered together vary from a few to several tens of μm. The single crystals of the hydrate have an isotropic spongelike crystalline structure with the mean size of the pores about several hundred of nm. Two granular samples of $CH_4$ hydrates with visually different grain sizes (2÷3 mm and ~1 mm for samples 1 and 2 respectively) were taken for thermal conductivity measurements. Before the measurement the $CH_4$ hydrates were kept at liquid nitrogen temperature. The low temperature part of the setup for measuring the thermal conductivity of $CH_4$ hydrates was described in.[5] The $CH_4$ hydrate was mounted in the measurement cell in the temperature range 80÷120 K. The hydrate powder in the cell was lightly compressed at pressures up to 0.70 MPa to increase its density. The cell was closed hermetically with a copper cover after the helium gas (1 bar) was filled into the cell. The volume density of the sample was $v_{sample}=V_{sample}/V_{cell}=0.65$ (where $V_{sample}$ and $V_{cell}$ are the volumes of the hydrate and cell, respectively). The main problem in measuring the thermal conductivity of the $CH_4$ hydrate sample is connected with the fact that $CH_4$ hydrate is highly porous. In such porous samples, the heat conduction is strongly restricted by the thermal contact resistance between the adjacent crystalline grains. In this study the effect of the thermal boundary resistance in the $CH_4$ hydrate sample was reduced by filling the measuring cell with He gas to provide a good thermal contact between the hydrate grains. The thermal conductivity of the gas is much lower than that of the grains but it can eliminate the influence of contact resistance.

The thermal conductivity of the methane hydrate was measured in the interval 2÷140 K using the steady-state technique. The effective thermal conductivity $k_{eff}$ of the sample can be presented as a sum of

$$k_{eff}(T) = k(T) \cdot v_{sample} + k_{He}(T)(1- v_{sample}), \qquad (1)$$

where $k(T)$ and $k_{He}(T)$ is the thermal conductivity of $CH_4$ hydrate and $^4He$ gas accordindly. Under a pressure of $100 \div 10^5$ Pa, $k_{He}$ is independent of pressure and is considerably lower than the thermal conductivity of $CH_4$ hydrate. The heat exchange gas eliminates (decreases) the temperature jump at the grain boundary because it participates in the heat transfer between the contacting grains. Eq. (1) relating the effective and true thermal conductivities was obtained assuming that the heat flow was parallel to the parallel layers of two substances.[6-8]

The $k_{eff}$ values measured on two different samples are shown in Fig. 1. The values of $k_{eff}$ of samples 1 and 2 agree within the experimental error (symbols ●, ○, respectively). The samples differed in the sizes of grains. $k_{eff}$ is strongly dependent on the compressing pressure inside the sample in the absence of the heat-transferring gas. The compressing pressure can be produced mechanically using a copper piston. When the compressing pressure increases, the area of the grain contact becomes larger. As a result, the thermal boundary resistance decreases. The thermal conductivity of sample 2 was measured as a function of temperature in three special tests with different thermal contact resistances inside the same sample. In test 1 the thermal contact resistance was the highest (the measuring cell was filled with $CH_4$ hydrate and neither heat-transferring gas, nor compressing pressure were used). In the test 2, the $CH_4$ hydrate was compressed mechanically with a copper piston to reduce the thermal contact resistance by increasing the grain contact areas. The sample contained no heat-transferring gas. In test 3, the heat-transferring gas was introduced into the sample and the compressing pressure was applied. The effective thermal conductivities are shown in Fig. 1. The lowest effective thermal conductivity (test 1) is mainly determined by the thermal contact resistance at the grain boundaries of a heterogeneous sample. In fig.1 the symbols (□) correspond mainly to the behavior of thermal resistance. The highest thermal conductivity in test 3 corresponds to a case when the contact resistance is not significant because of the heat-transferring gas (symbols ○). At low temperatures the temperature dependence of the thermal conductivity $k_{eff}(T)$ of $CH_4$ hydrate changes monotonically with lowering temperature. There are no features in the curve $k_{eff}(T)$ that might suggest possible intensive adsorption of helium by the sample or formation of superfluid helium film at the grain boundary or formation of He-hydrate.

## 3. Results and analysis

The thermal conductivity $k(T)$ of the $CH_4$ hydrate calculated from experimental data $k_{eff}$ by Eq. (1) is shown in Fig. 2 ($v_{sample}$= 65%). The estimated measured thermal conductivity of $CH_4$ hydrate is in good agreement with previous isolated measurements.[9-11] Above $T = 80$ K the experimental results on thermal conductivity can be approximated to a good accuracy by the linear dependence $k(T) = 0.43 + 0.00021 T$ J/(m K).

The thermal conductivity of $CH_4$ hydrate coincides within the experimental error with the thermal conductivity for dioxolane hydrate.[12] This fact is quite surprising because these are cubic but have different structures (type I space group $Pm3n$ in $CH_4$ hydrate and type II space group $Fd3m$ in dioxolane hydrate) and, in particular, different guest molecules. This result may originate because the thermal conductivity of the hydrate is basically determined both by the interaction between the phonons and low frequency local vibrations of the guest molecules and by proton disorder in clathrate-hydrate compounds. The similar behavior of $k(T)$ of $CH_4$ hydrate, in which the rotation of the guest molecules is weakly damped,[15] and dioxolane hydrate, in which rotational motion is strongly hindered, may indicate that the contribution of the phonon-rotation interactions to the thermal conductivity of the gas hydrate is of little importance.

In addition to the data for the hydrates, Fig. 2 illustrates the thermal conductivity of sample 3. This sample is heterogeneous ice formed after decomposition of sample 2 ($CH_4$ hydrate) at $T = 230 \div 240$ K. In this temperature region the separation of the hydrate into $CH_4$ and solid ice is the fastest.[16] As a result, polycrystalline Ih – type ice is formed, which however does not destroy the initial granular structure of the sample. The density of packing in the measuring cell decreases insignificantly ($v_{sample}= 63\%$). The thermal contact resistance in sample 3 was reduced by the same procedure as for the $CH_4$ hydrate samples – by filling He gas into the cell. The measured effective thermal conductivity of the sample 3 is shown in Fig. 1. The thermal conductivity estimated by Eq. (1) is shown in Fig. 2. The thermal conductivity of the granular ice is considerably higher than that of the $CH_4$ hydrate. The dependence $k(T)$ of the thermal conductivity of granular ice is typical of heavily deformed fine-grain crystal dielectric. At the lowest temperatures of the experiment, the dependence $k(T) \sim T^3$ is observed, which typically occurs at phonon scattering at the crystal grain boundaries. The measurement of the thermal conductivity of ice proves that the proposed technique of measuring the thermal conductivity of porous substances at low temperatures is completely reliable. This measurement has also provided a test of reliability of the experimental data.

The thermal conductivity is analyzed within the Debye model. The thermal conductivity is calculated by

$$k(T) = \frac{1}{2\pi^2 v k_B T^2} \int_0^{\omega_D} \tau_{total} \frac{\hbar \omega^4 e^{\hbar\omega/k_B T}}{(1-e^{\hbar\omega/k_B T})^2} d\omega, \qquad (2)$$

where $\omega_D$ is the Debye frequency and **v** is the phonon group velocity. $\tau_{total}$ is the characteristic relaxation time due to different processes of phonon scattering: umklapp processes ($\tau_U$), point defect scattering $\tau_D$, boundary scattering ($\tau_B$) and the scattering leading to the glasslike thermal conductivity ($\tau_G$).

$$\tau^{-1}_{total} = \tau^{-1}_U + \tau^{-1}_B + \tau^{-1}_D + \tau^{-1}_G \quad (3)$$

$\tau_U^{-1} = B\,\omega^2\,T\,exp(-b/T)$ for the umklapp process.

$\tau_B^{-1} = \mathbf{v}/L$ for the boundary scattering. $L$ is the size of crystal grains.

$\tau^{-1}_D = R\,\omega^4$ for the Rayleigh scattering by point defects.

There are two different approaches to describe $\tau^{-1}_G$. It was taken into account, that $\tau_{total}$ was no less than $\tau_{min}=2/\omega$. At high temperature the localized excitations in gas hydrate are fully activated and the mean free path for phonon scattering is therefore nearly a constant.

In the analysis[12] of the glasslike thermal conductivity of 1,3-dioxolane clathrate hydrate the dominant phonon scattering was attributed to the tunneling state arising from the disorder in the hydrate. Ice and gas hydrates have practically similar proton disorder, therefore this cannot be the reason for the low thermal conductivity.

The gas hydrates have two different types of disorder: proton (hydrogen) disorder in the host lattice and orientation disorder of the guest molecules in the hydrate cages. The phenomenological soft-potential model[17, 18] may be used to calculated $\tau_G^{-1}$.

$$\tau^{-1}_G = A\omega\tanh\left(\frac{\hbar\omega}{2k_B T}\right) + A\omega\left[\frac{T}{W}\right]^{3/4} + \frac{A\omega}{6\sqrt{2}}\left[\frac{\hbar\omega}{W}\right]^3, \quad (4)$$

A and W are the model parameters. There are three contributions of phonon scattering from tunneling states, classical relaxors and soft quasiharmonic vibrations.

The strong interaction between rotational and translation motions of the guest molecules with lattice vibrations leads to the glasslike behavior of the thermal conductivity of gas hydrates.[1, 19] The dominant resonant phonon scattering was accounted for by the strong coupling of acoustic phonon modes with guest vibrations. The strong couplings result from avoiding crossing of the enclathrated guest localized modes with the branches of the host lattice acoustic phonons. In the phenomenological resonant-scattering model[1]

$$\tau^{-1}_G = D\,\frac{\omega^2}{(\omega^2 - \omega_0^2)^2} \quad (5)$$

where ***D*** is the scattering constant dependent on the number of guest molecules and on the coefficient relating to the strength of the water lattice-guest coupling, and $\omega_0$ is the vibrational frequency of the guest responsible for resonant scattering. In fig. 3 the results of the fits by Eqs. (4) or (5) are shown as solid curves. In the fits for the $CH_4$ hydrate and ice powder we used the parameters of Table 1. The phenomenological soft-potential model describes the experimental data very well and the resonant-scattering model gives a good description only below 25 K. The thermal conductivity for the powder ice is in good agreement with the calculation taking into account the boundary and point defect strong scattering. In $CH_4$ hydrate the rotational motion of $CH_4$ at low temperature is almost free, like the quantum rotor in rare gas solids or solid methane. The phonon resonant scattering by quantum $CH_4$ rotors does not lead to glass-like thermal conductivity.[22]

The resonant-scattering model does not take into account the velocity dispersion possibly caused by resonant scattering. The intensive dispersion of the sound velocity implies that the acoustic phonons whose energy is equal to or higher than the energy of the local mode propagate in the diffusion condition. Their group velocity is close to zero. A similar picture is observed in structural glasses where high-frequency acoustic phonons experience strong resonance scattering by boson-peek excitation.[23]

The conductivities of semiconductor cage (type I) structures[24-26] and hydrates are similar. The reason for this similarity may be in the identical cage structures. The clathrate framework structure is more open as compared to ordinary ice. It consists of water molecules. Since water has flexible H-bonds, the length of the hydrogen bonds and the spatial angles between them appear to be distorted.

$CH_4$ hydrate has a complex cubic crystal structure of type I. The unit cell of the structure (*Pm*3*n*, *a* = 12 Å) contains 46 water molecules comprising two 12-hedron and six 14-hedron cavities, each of which is occupied by a methane molecule. The interaction of the lattice vibrations and the local optical mode, which stabilizes the structure of the gas hydrate can induce new types of excitations, e. g., two-level tunnel systems. On the other hand, the high-frequency phonons can transfer the heat in the diffusion condition (the Ioffe – Regel limit). These two factors typical of ordinary structural glasses can also be used to explain the glass-like behavior of the thermal conductivity of $CH_4$ hydrate having this soft of loose structure.

## 4. Conclusions

The thermal conductivity of synthesized methane hydrate has been measured for the first time in the temperature region 2÷140 K. The shape of ***k***(*T*) is typical of the thermal conductivity of amorphous solids. It is shown that after separation of the hydrate into ice and methane, at 240 K, the thermal conductivity of the ice exhibits the dependence $k(T)$ typical of heavily deformed fine-grain polycrystals.

## Acknowledgments

This work has been partially funded as a project by Forschungszentrum Jülich.

## 5. References

1. J. S. Tse, Mary Anne White. *J.Phys.Chem*., **92**, 5006 (1988).
2. J. Baumert, C. Gutt, V.P. Shpakov, J.S. Tse, M. Krisch, M. Muller, H. Requardt, D.D. Klug, S. Janssen, and W. Press. *Phys. Rev.* **B 68**, 174301 (2003).
3. V. F. Sears, B. M. Powell, J. S. Tse, C. I. Ratcliffe and Y. P. Handa. *Physica* **B 180**, Part 2, 658 (1992).
4. D. Staykova, Th. Hansen, A. N. Salamatin, and W. F. Kuhs.; Proc. Fourth Intern. Conf. on Gas Hydrates, Yokohama, May 19-23, 2002, pp. 537-542.
5. A. I. Krivchikov, B. Ya. Gorodilov, O. A. Korolyuk, to be published.
6. Missenard A. Conductivite Thermique des Solides, Liuides, Gaz et de Leurs Melanges. Editions Eyrolles, Paris, 1965. (In French)
7. Dulnev, G.N., Zarichnyak, Yu.P. Heat Conductivity of Mixtures and Composite Materials. Energiya, Moscow, 1974. (In Russian).
8. Litovsky Ye. Ya., Puchkelevich, N. A. Thermal Properties of Heat-resistant Materials. Handbook. Metallurgiya, Moscow, 1982. (In Russian).
9. J. G. Cook, D. G. Leaist. *Geophys. Res. Lett*., **10**, 397 (1983).
10. Davidson, D., Gas hydrates as clathrate ices, in Cox, J., ed., Natural gas hydrates-- properties, occurrence and recovery: Butterworth, Woburn, MA, p.1-16, 1983.
11. Hsing Wang, unpublished.
12. N. Ahmad, W. A. Phillips. *Solid State Comm*., **63**, 167 (1987).
13. G. A. Slack. *Phys. Rev*. **B**, 3065 (1980).
14. J. Klinger and G. J. Rochas. *Phys. Chem*, **87**, 4155 (1983).
15. C. Gutt, B. Asmussen, W. Press, C. Merkl, H. Casalta, J. Greinert, G. Bohrmann, J. S. Tse and A. Hüller. *Europhys. Lett*., **48**, 269 (1999).

16. Laura A. Stern, Susan Circone, Stephen H. Kirby, and William B. *J.Phys.Chem*. **B 105**, 1756 (2001).
17. U. Buchenau, Yu. M. Galperin, V. L. Gurevich, D. A. Parchin, M. A. Ramos, and H. R. Schober. *Phys. Rev*. **B 46**, 2798 (1992).
18. M. A. Ramos, U. Buchenau. *Phys. Rev*. **B 55**, 5749 (1997).
19. J. S. Tse, V. P. Shpakov, V. V. Murashov, V. R. Belosludov. *J. Chem. Phys*., **107**, 9271 (1997).
20. S. Sasaki, T. Kumazaki, I Suwa, T. Kume and H. Shimizu. *J. Phys.: Condens. Matter* **14**, 10445 (2002).
21. V. F. Petrenko, and R. W. Whithworth, Physics of ice, Oxford university press, 1999.
22. A. I. Krivchikov, B. Ya. Gorodilov, V. G. Manzhelii and V. V. Dudkin, *Low Temp. Phys*. **29**, (2003).
23. E. Courtens, M. Foret, B. Hehlen, B. Ruffle, R. Vacher. *J. Phys.: Condens. Matter* **15**, S1281 (2003).
24. J. L. Cohn, G. S. Nolas, V. Fessatidis, T. H. Metcalf, and G. A. Slack. *Phys. Rev. Lett*. **82**, 779 (1999).
25. G. S. Nolas, T. J. R. Weakley, J. L. Cohn, and R. Sharma. *Phys. Rev*. **B 61**, 3845 (2000).
26. A. Bentien, M. Christensen, J. D. Bryan, A. Sanchez, S. Paschen, F. Steglich, G. D. Stucky, and B. B. Iversen. *Phys. Rev*. **B 69**, 045107 (2004).

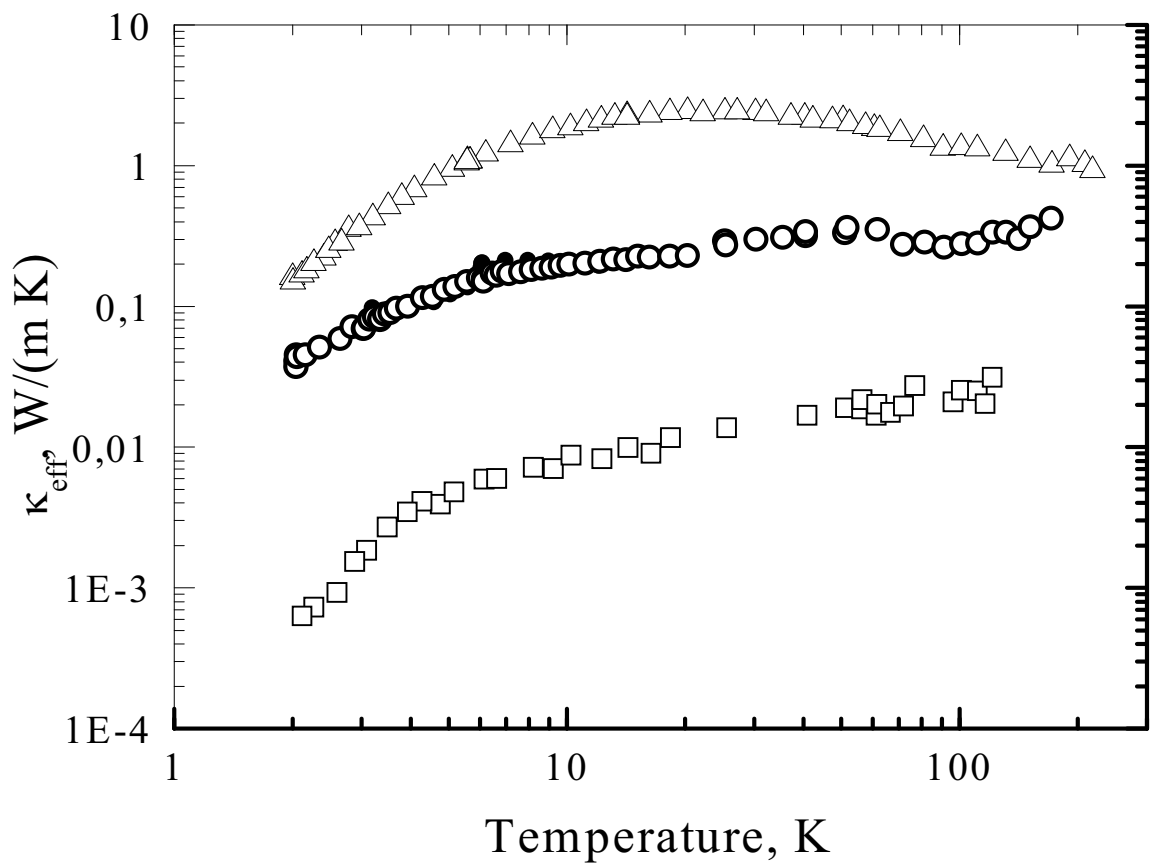

Fig. 1. The effective thermal conductivity of grainy methane hydrate: the samples 1- ●, 2-○ - with heat-transferring gas $^4He$ and □ the sample 2 without heat-transferring gas. For comparison we show the data for sample 3 (Δ) – granular ice with heat-transferring gas $^4He$. The ice was prepared from hydrate sample 2.

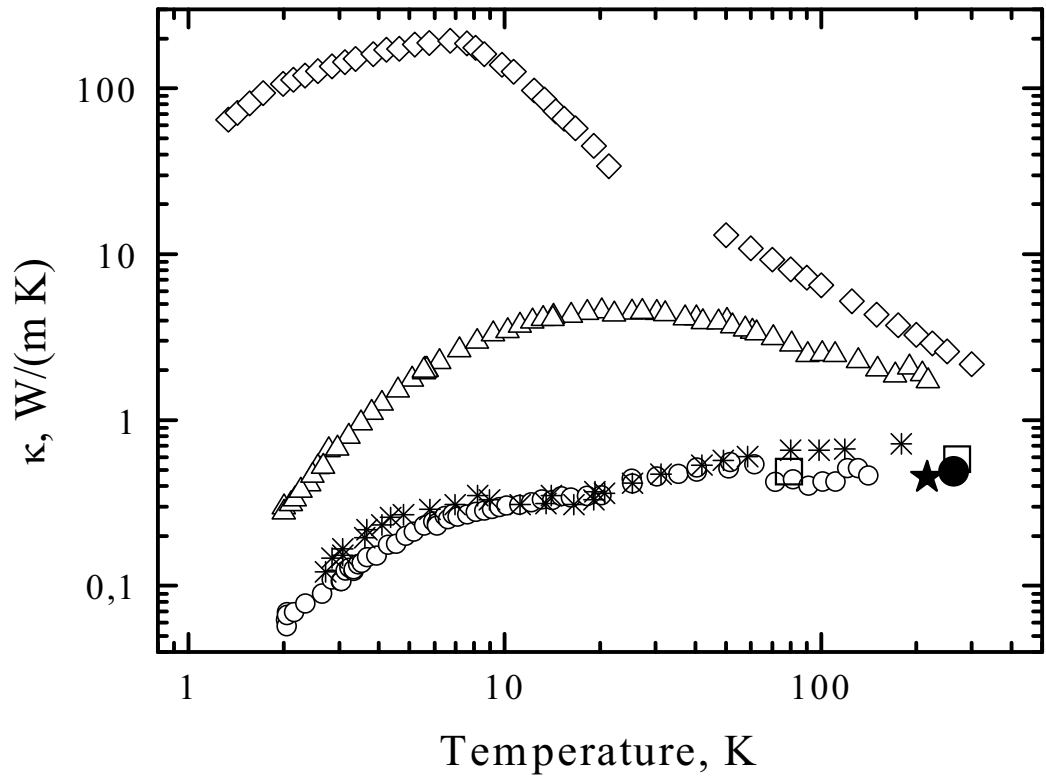

Fig. 2. Thermal conductivity of clathrate hydrates and ice: $CH_4$ hydrate ○ - our data; ★ - ref. 9, ● - ref. 10, □ - ref. 11, *- dioxolane hydrate [12]; ∆- powder ice after $CH_4$ hydrate decomposition our data (sample 2); ◊- ice [13,14] for bulk ice.

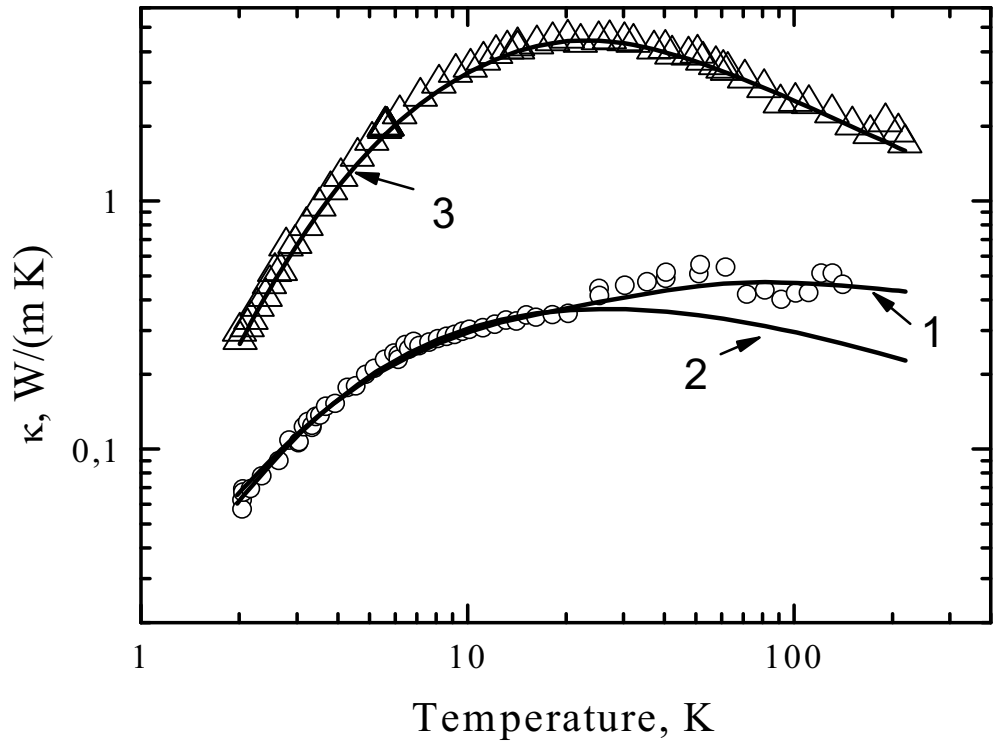

Fig. 3. Thermal conductivity of clathrate hydrates and ice experimental data: o - $CH_4$ hydrate; Δ- powder ice after $CH_4$ hydrate decomposition and calculated curves by the Debye theory: 1- soft-potential model; 2- resonant-scattering model and 3 – the model for *k*(*T*) considering boundary and point defect strong scatterings.

Table 1. Parameters to fit the thermal conductivity of $CH_4$ hydrate and ice. For the hydrate $\Theta_D = 218$ K and v = 2100 m/s.[20] For the ice $\Theta_D = 226$ K and v = 2640 m/s.[21]

| $B \cdot 10^{27}$, s K$^{-1}$ | $b$, K | $R \cdot 10^{48}$, s$^3$ | $L \cdot 10^6$, m | $A \cdot 10^{19}$ | $W$, K | $D \cdot 10^{38}$, s$^{-3}$ | $\omega_0$, K |
|---|---|---|---|---|---|---|---|
| 4.8 | 45 | 2.0 | 3.74 | 3.1 | 6.5 | 5.1 | 150 |